# Doping-induced structural transformation in the spin-1/2 triangular-lattice antiferromagnet $Na_2Ba_{1-x}Sr_xCo(PO_4)_2$

Chuandi Zhang,[1, *] Qianhui Xu,[1, *] Xu-Tao Zeng,[1, *] Chao Lyu,[2] Zhengwang Lin,[1] Jiazheng Hao,[3] Sihao Deng,[3] Lunhua He,[3, 4, 5, †] Yinguo Xiao,[6] Yu Ye,[2] Ziyu Chen,[1] Xian-Lei Sheng,[1] and Wentao Jin[1, ‡]

[1]*School of Physics, Beihang University, Beijing 100191, China*

[2] *State Key Laboratory for Mesoscopic Physics and Frontiers Science Center for Nano-optoelectronics, School of Physics, Peking University, Beijing 100871, China*

[3]*Spallation Neutron Source Science Center, Dongguan 523803, China*

[4]*Beijing National Laboratory for Condensed Matter Physics, Institute of Physics, Chinese Academy of Sciences, Beijing 100190, China*

[5]*Songshan Lake Materials Laboratory, Dongguan 523808, China*

[6]*School of Advanced Materials, Peking University Shenzhen Graduate School, Shenzhen 518055, China*

The effects of Sr doping on the structural properties of $Na_2BaCo(PO_4)_2$, a spin-1/2 triangular-lattice antiferromagnet as a quantum spin liquid candidate, are investigated by complementary x-ray and neutron powder diffraction measurements. It is found that in $Na_2Ba_{1-x}Sr_xCo(PO_4)_2$ (NBSCPO), the trigonal phase (space group $P\bar{3}m1$) with a perfect triangular lattice of $Co^{2+}$ ions is structurally stable when the doping level of Sr is below 30% ($x \leq 0.3$), while a pure monoclinic phase (space group $P2_1/a$) with slight rotations of $CoO_6$ octahedra and displacements of $Ba^{2+}/Sr^{2+}$ ions will be established when the Sr doping level is above 60% ($x \geq 0.6$). Such a doping-induced structural transformation in NBSCPO is supported by first-principles calculations and Raman spectroscopy. $Na_2SrCo(PO_4)_2$, a novel spin-1/2 antiferromagnet with glaserite-type structure, although monoclinically distorted, exhibits no long-range magnetic order down to 2 K and a similar negative Curie-Weiss temperature as $Na_2BaCo(PO_4)_2$ with a perfect triangular lattice, suggesting the robustness of magnetic exchange interaction against the Ba/Sr substitutions.

## INTRODUCTION

Spin-1/2 triangular-lattice antiferromagnets have attracted much attention recently. The combination of strong geometrical frustration and quantum fluctuations leads to a variety of exotic quantum states in these materials, such as quantum spin liquids, [1–4] intrinsic quantum Ising magnets, [5–11] quantum spin state transitions, [12–16] and so on. Among them, the quantum spin liquid (QSL) state proposed by P. W. Anderson in 1973, [17] in which the spins remain liquid-like without showing any long-range magnetic order even at absolute zero temperature, has been intensively studied in past decades. A QSL state can host long-range spin entanglements, topological orders, and fractional spin excitations known as spinons, which holds great potential for quantum communication and computation.

Numerous experimental efforts have been devoted to the explorations of QSL materials in spin-1/2 quantum magnets with antiferromagnetic interactions in a triangular network of magnetic ions. [18] For example, the triangular-lattice Yb-based antiferromagnets $YbMgGaO_4$, $NaYbO_2$, $NaYbS_2$, and $NaYbSe_2$ with effective spin-1/2 for $Yb^{3+}$ ions have been proposed as QSL candidates. [19–25] However, confirming a genuine QSL state in these compounds is challenging, as the existence of either structural disorder or anisotropic spin interactions in them usually results in a spin glass or magnetically disordered ground state, which is difficult to be distinguished from the QSL state.

Very recently, $Na_2BaCo(PO_4)_2$ (NBCPO) crystallizing in a trigonal structure (space group $P\bar{3}m1$), in which the $Co^{2+}$ ions (Kramers ions with electronic configuration $3d^7$ and effective $S = 1/2$) reside on a perfect triangular lattice, has drawn a great deal of interests. [16,26–28] As reported by Zhong *et al*. and Lee *et al*., no long-range magnetic order was observed in NBCPO by neutron diffraction and muon spin relaxation measurements, down to 300 mK and 80 mK, respectively. [26,27] In addition, inelastic neutron scattering and thermal transport experiments on NBCPO revealed a low-energy magnetic continuum indicative of spinon excitations and a nonzero residual thermal conductivity, respectively, supporting a QSL-like state. [16,26] However, the heat capacity measurement by Li *et al*. unraveled a weak antiferromagnetic order at $T_N \sim$ 148mK in NBCPO. [16] So far, it is still controversial about the exact magnetic ground state of NBCPO, necessitating further experimental investigations on this novel material and its structural derivatives.

Compared with the Yb-based triangular-lattice antiferromagnets, NBCPO is believed to possess no structural disorder and dominant isotropic exchange interactions, so that it is considered to be in close proximity to an $S = 1/2$ triangular Heisenberg antiferromagnet, the first conceptual model system historically proposed to realize the QSL state. As the magnetic interactions strongly depend on the atomic distances between magnetic ions, structural modifications to NBCPO by means of chemical substitutions or applying physical pressure will provide a way for tuning its magnetic properties and understanding the correlations between the structure and the magnetism.

In this paper, we have performed a fine tuning of the crystal structure of NBCPO by systematic chemical substitution of Sr for Ba. Using solid state reaction method, polycrystalline samples of $Na_2Ba_{1-x}Sr_xCo(PO_4)_2$ (NBSCPO) were synthesized. Detailed structural characterizations of the NBSCPO samples were carried out by complementary x-ray and neutron powder diffraction measurements. Rietveld refinements to the diffraction patterns revealed that the trigonal structure is stable at low Sr doping levels ($x \leq 0.3$), and a pure monoclinic phase with slight rotations of $CoO_6$ octahedra and displacements of $Ba^{2+}/Sr^{2+}$ ions is formed at high Sr doping levels ($x \geq 0.6$). The lattice constants and the atomic distances between the $Co^{2+}$ ions shrink with increasing Sr doping in both phase regimes. At intermediate Sr doping levels ($0.4 \leq x \leq 0.5$), the trigonal and monoclinic phases coexist. The doping-induced structural transformation in NBSCPO is supported by first-principles calculations and Raman spectroscopy. The dc magnetic susceptibility measurements reveal the absence of long-range magnetic order down to 2 K in $Na_2SrCo(PO_4)_2$ (NSCPO), which is a novel monoclinic-distorted spin-1/2 antiferromagnet with less geometrical frustration, and the robustness of magnetic exchange interaction against the Ba/Sr substitutions.

**EXPERIMENTAL DETAILS AND CALCULATION METHODS**

Polycrystalline samples of $Na_2Ba_{1-x}Sr_xCo(PO_4)_2$ ($x$ = 0, 0.2, 0.3, 0.4, 0.5, 0.6, 0.8, and 1) were synthesized by the solid-state reaction method. A stoichiometric mixture of dried $Na_2CO_3$ (99.99%), $BaCO_3$ (99.95%), $SrCO_3$ (99.95%), CoO (99%), and $(NH_4)_2HPO_4$ (99.99%) was ground well together with the catalyst $NH_4Cl$ (99.99%) in the molar ratio of 2:1, pressed into pellets, sintered in air at 700 °C for 24 h in alumina crucibles, and finally cooled down to room temperature. The sintering process was repeated for three times to minimize the impurity phases.

Structural characterizations of the polycrystalline samples of NBSCPO were performed using both x-ray diffraction (XRD) and neutron powder diffraction (NPD) measurements. The XRD patterns on all samples were collected using a Bruker D8 diffractometer in Bragg-Brentano geometry with Cu-K$\alpha$ radiation ($\lambda$ = 1.5406 Å) at room temperature in a range of 10-100°. The NPD experiments were carried out on the time-of-flight (TOF) diffractometer GPPD (General Purpose Powder Diffractometer) at China Spallation Neutron Source (CSNS), Dongguan, China. [29] Samples with the mass around 1.6-1.8 g were loaded into vanadium containers, and the neutron diffraction patterns were collected at room temperature with wavelength band from 0.1 to 4.9 Å. The program FullProf was used for the Rietveld refinement of the crystal structures of all compounds.[30]

The Raman spectra of NBCPO and NSCPO were collected under ambient conditions using a commercial confocal Raman microscope, alpha 300R WITec system (WITec GmbH, Ulm, Germany). A continuous-wave 532 nm laser was utilized as the excitation, and a 100× short-working-distance objective lens with a numerical aperture (NA) of 0.95 was used to focus the laser and collect signals.

Magnetic properties of polycrystalline $Na_2BaCo(PO_4)_2$ ($x$=0) and $Na_2SrCo(PO_4)_2$ ($x$ = 1) were characterized using a Quantum Design Magnetic Property Measurement System (MPMS). The temperature dependence of the dc magnetic susceptibility was measured during warming from 2 K to 300 K in an applied magnetic field of 1000 Oe under the zero-field-cooling (ZFC) condition.

First-principles calculations have been carried out based on the density-functional theory (DFT) as implemented in the Vienna *ab initio* simulation package (VASP) [31,32], using the projector augmented wave method [33] and Perdew-Burke-Ernzerhof (PBE) exchange-correlation functional approach [34]. The plane-wave cutoff energy was set to 320 eV. The Monkhorst-Pack $k$-point mesh [35] of size $4 \times 6 \times 2$ was used for the BZ sampling.

The crystal electric field (CEF) of NBCPO and NSCPO were calculated by constructing and solving a point-charge model with the information of ligand environment as implemented in *PyCrystalField* package [36] and the structural parameters determined experimentally.

**RESULTS AND DISCUSSIONS**

The room-temperature XRD and NPD patterns of $Na_2Ba_{1-x}Sr_xCo(PO_4)_2$ (x = 0, 0.2, 0.3, 0.4, 0.5, 0.6, 0.8, and 1) are showed in Fig. 1. At low Sr doping levels ($x \leq 0.3$), all the diffraction peaks can be well indexed using the trigonal structure as reported for undoped NBCPO in Ref. 26, showing only the peak shifting associated with the lattice contraction due to substitutions of $Sr^{2+}$ (ionic radius 1.36 Å) for $Ba^{2+}$ (ionic radius 1.61 Å). At high Sr doping levels ($x \geq 0.6$), numbers of reflections split and additional weak reflections emerge, suggesting a symmetry lowering from trigonal to monoclinic, as supported by the Rietveld refinements represented below. For instance, as shown in Fig. 1(b) and Fig. 1(d), the $(110)_t$ reflection and its equivalents in trigonal notation splits into the $(310)_m$ and $(020)_m$ reflections in monoclinic notation, and additional weak reflections marked by asterisks including $(203)_m$, $(113)_m$ and $(311)_m$ appear. The patterns for $x \geq 0.6$ fit well with a structure with a monoclinic distortion together with a tiny amount of $Sr_2P_2O_5$ impurity formed in the sintering process. The appearance of the weak reflections indexed with $(H\ K\ odd)_m$ indicates the doubling of lattice constant $c$ related to the trigonal-monoclinic transformation, which will be discussed in details later. At intermediate Sr doping levels ($0.4 \leq x \leq 0.5$), the trigonal phase can still be identified in the diffraction patterns, in addition to the emergent monoclinic phase, giving rise to a coexistence of two structural phases in these compositions.

The XRD and NPD patterns of all samples collected at room temperature are simultaneously refined using the Rietveld refinement program FullProf suite. Fig. 2 shows the refinement results of $Na_2Ba_{1-x}Sr_xCo(PO_4)_2$ for $x = 0$ (a and b), 0.3 (c and d), 0.6 (e and f) and 1 (g and h). The Sr doping-induced structural transformation from trigonal to monoclinic symmetry is clearly visible, as shown in the insets, evidenced by the splitting of a single $(110)_t$ reflection (a-d) into $(310)_m$ and $(020)_m$ reflections (e-f) as well as the emergence of a nearby $(311)_m$ reflection (e-f). Note that the asymmetric peak profiles in the XRD patterns is due to the combined contributions from both Cu-$K\alpha_1$ and Cu-$K\alpha_2$ radiations with the intensity ratio of 2:1, leading to a doublet of peaks for the $(110)_t$ reflection (a and c) and quartet of peaks for the $(310)_m/(020)_m$ reflection pair (e and g).

The Sr-doping induced structural transformation from trigonal to monoclinic symmetry is illustrated in Fig. 3. All samples of $Na_2Ba_{1-x}Sr_xCo(PO_4)_2$ are found to adopt the glaserite-type $XY_2M(TO_4)_2$ structure [37] in which the Na, Ba/Sr, Co and P sites are co-ordinated by 10, 12/10, 6 and 4 $O^{2-}$ ions, respectively. The structure of a pure NBCPO ($x = 0$) determined here by x-ray and neutron powder diffraction is very similar to that determined by single-crystal XRD in Ref. 26, with the $CoO_6$ octahedra forming a perfect triangular network in the $ab$ plane (see Fig. 3(a)). These magnetic layers are separated by a single layer of nonmagnetic $BaO_{12}$ polyhedral, with the $PO_4$ tetrahedra and $Na^+$ ions filling the remaining space. As is shown in Fig. 3(c) viewed from the c axis, each $CoO_6$ octahedron is linked to six undistorted corner-shared $PO_4$ tetrahedra, showing a threefold rotational symmetry in accord with the $P\bar{3}m1$ space group. In addition, the $Ba^{2+}$ ions locate exactly at the center position of neighboring $Co^{2+}$ ions along the c direction, yielding a 180° Co-Ba-Co angle.

The introduction of Sr dopants, however, tends to induce some lattice distortions, due to the considerable difference of ionic radius between $Sr^{2+}$ ($r$ = 1.36 Å) and $Ba^{2+}$ ($r$ = 1.61 Å). As a response, the $CoO_6$ octahedra in neighboring layers rotate slightly in opposite directions, which doubles the lattice constant $c$, as illustrated in Fig. 3(b) for a pure NSCPO ($x$ = 1). The slight rotation of the $CoO_6$ octahedra also distorted the $PO_4$ tetrahedra linked with them (Fig. 3(d). Furthermore, the $Sr^{2+}$ ions show some displacements in the $ab$ plane, away from the center position of neighboring $Co^{2+}$ ions on adjacent magnetic layers. The polyhedral distortions and ionic displacements induced by Sr doping break the threefold rotational symmetry, leading to a symmetry lowering from trigonal to monoclinic.

Such a trigonal-monoclinic structural transformation involves the change of the unit cell. As shown in Fig. 3, associated with the symmetry lowering, the in-plane and out-of-plane lattice constants transform into $a_m \approx \sqrt{3}a_t$, $b_m \approx b_t$, and $c_m \approx 2c_t$, respectively, with the subscript m and t denoting the monoclinic and trigonal notation. This unit cell conversion and the symmetry lowering account for the splitting of overlapped equivalent $(110)t/(\bar{1}20)t$ reflection into non-equivalent $(020)m$ and $(310)m$ upon Sr doping, as well as the absence of correspondence for $(H\ K\ odd)m$ reflection in trigonal symmetry, considering the conversion rule of $Hm = Kt - Ht$, $Km = Kt + Ht$, and $Lm = 2Lt$.

In Table 1, we have listed the structural parameters of NSCPO ($x$ = 1) determined by simultaneous refinements of both the x-ray and neutron diffraction patterns. This compound, which was never reported before, crystallizes in the monoclinic symmetry with the space group of $P2_1/a$. We note that this glaserite-type structure with the same space group was ever reported for $Na_2SrMg(PO_4)_2$, which contains nonmagnetic $3d$ $Mg^{2+}$ ions.

[38] Compared with NBCPO, the unit cell of NSCPO is four times large in volume. As a result of monoclinic distortion, the unique Co site in a trigonal lattice evolves into two crystallographically nonequivalent sites, denoted as $Co_1$ (2a) and $Co_2$ (2d), respectively, with their surrounding $CoO_6$ octahedra rotating in opposite directions.

The evolution of room-temperature lattice parameters with Sr doping is shown in Fig. 4(a) and 4(b). Both the in-plane and out-of-plane lattice parameters are found to decrease with the introduction of Sr dopants. This is consistent with the fact that the $Sr^{2+}$ ions are considerably smaller than the $Ba^{2+}$ ions. Such a lattice contraction will also lead to a decrease of both interlayer and intralayer Co-Co distances and accordingly that of the magnetic exchange interactions. Furthermore, as stated above, the substitution of $Sr^{2+}$ for $Ba^{2+}$ distorts the unit cell from the trigonal symmetry. In the monoclinic regime, along the $c$ axis, the $Co_1$-Ba/Sr-$Co_2$ bond angle gradually deviates from 180°, as shown in Fig. 4(b), finally reaching 166.3° for the pure NSCPO ($x = 1$), which suggests a sizeable in-plane displacement of Ba/Sr ions associated with the trigonal-monoclinic structural transformation. In addition, the in-plane $Co_1$-$Co_1$-$Co_1$ bond angle shrinks continuously from 60° for a perfect triangular lattice when entering the monoclinic regime. Such a doping-induced distortion of the triangular network of $Co^{2+}$ ions (with the $Co_1$-$Co_1$-$Co_1$ angle of 59.54° for NSCPO), although slight, will help to release some geometrical frustration and might result in a different magnetic ground state.

The symmetry lowering in NSCPO compared with NBCPO is also evidenced by the Raman spectroscopy. According to the factor-group analysis for the space group of $P\bar{3}m1$ (trigonal symmetry) performed using the program SAM from the Bilbao Crystallographic Server [39], 11 Raman active modes in total are allowed for NBCPO.

Especially, four normal phonon modes of the $PO_4$ tetrahedra at high wavenumbers (> 400 cm$^{-1}$) are expected for such phosphates with glaserite-type structure [40], including the symmetric stretching vibration mode around 1000 cm$^{-1}$, symmetric bending vibration mode around 440 cm$^{-1}$, asymmetric stretching vibration mode around 1120 cm$^{-1}$, and asymmetric bending vibration mode around 590 cm$^{-1}$, well consistent with the positions of Raman peaks in the range of 400-1200 cm$^{-1}$ observed for NBCPO as shown in Fig. 5. However, for NSCPO, these four peaks marked by 1-4 become clearly broadened and asymmetric, suggesting the emergence of multiple modes with close energies that can not be resolved within the instrument resolution. Such a peak broadening for NSCPO is fully in accord with the fact that the $PO_4$ tetrahedra become distorted due to slight rotations of $CoO_6$ octahedra and the corresponding internal P-O bond lengths become non-equivalent in the monoclinic phase, thus further corroborating the symmetry lowering in NSCPO compared with NBCPO.

Furthermore, the structural stabilities of trigonal NBCPO and monoclinic NSCPO are checked by performing first-principles calculations. The experimentally determined lattice constants of both compounds were adopted for single-point energy calculations. Based on lattice relaxation calculations, the lattice constants are optimized to $a$ = 9.606 Å, $b$ = 5.432 Å, $c$ = 13.707 Å, $\alpha = \gamma = 90°$, $\beta = 89.751°$ for a possible monoclinic phase of NBCPO and $a$ = 5.280 Å, $c$ = 6.765 Å for a possible trigonal phase of NSCPO. By directly comparing the free energy of the monoclinic unit cell and the $\sqrt{3} \times 1 \times 2$ trigonal supercell with comparable volumes, our DFT calculations confirm that the trigonal phase of NBCPO and the monoclinic phase of NSCPO are structurally more stable, as shown in Table II. This is well consistent with our experiment observations that

NBSCPO undergoes a Sr doping-induced structural transformation from trigonal to monoclinic symmetry.

In order to check the effect of such a structural transformation on the magnetic properties, we have analyzed the alignment of energy levels of Co$^{2+}$ ions for both NBCPO and NSCPO. According to the CEF theory [41], both the CEF effect $\mathbf{H_{CEF}} = \sum_{n,m} B_n^m O_n^m$ and the spin-orbit coupling (SOC) $\mathbf{H_{SOC}} = \tilde{\lambda}\mathbf{S}\cdot\mathbf{L}$ should be considered in the Hamiltonian for 3d transition-metal ions, as they are of a similar magnitude. Here $O_n^m$ are the Stevens operators, $B_n^m$ are multiplicative factors called CEF parameters, n is the operator degree constrained by time-reversal symmetry to be even, m is the operator order with $-n \leq m \leq n$, $\mathbf{S}$ is the spin angular momentum, and $\mathbf{L}$ is orbital angular momentum, respectively. For Co$^{2+}$ ions, it is necessary to treat both the CEF and SOC non-perturbatively in the so-called intermediate coupling scheme, in which $\mathbf{H_{CEF}}$ acts only on $\mathbf{L}$. For free Co$^{2+}$ ions with the electronic configuration of 3$d^7$, the ground state is $^4$F, which splits into three energy levels ($\Gamma_2$, $\Gamma_5$, and $\Gamma_4$) in a predominantly cubic-symmetry crystal field, as shown in Fig. 6. The $\Gamma_2$ state can be represented by the fictitious orbital angular momentum $\tilde{\mathbf{L}} = 1$ and the spin angular momentum $\mathbf{S} = 3/2$. Under the action of SOC, the $(2\tilde{\mathbf{L}}+1)(2\mathbf{S}+1) = 3(2\mathbf{S}+1)$ manifold of states splits into a series of energy levels characterized by an effective quantum number $\tilde{\mathbf{J}} = 1/2$, 3/2 and 5/2. For NBCPO with trigonal symmetry, the trigonal distortion naturally splits the $\tilde{\mathbf{J}} = 3/2$ states of Co$^{2+}$ ions into two well-separated energy levels. For NSCPO with monoclinic symmetry, the cubic-symmetry crystal field is fully distorted into a lower-symmetry centrosymmetric field acting on the two non-equivalent Co$^{2+}$ sites (Co$_1$ and Co$_2$), as a consequence, enhancing the CEF splitting as well. Utilizing a point-charge model, the CEF energy levels can be calculated (see Table III for the CEF parameters). The atomic SOC strength of Co is taken as the effective SOC constant $\tilde{\lambda}$ of $Co^{2+}$ ions [42]. Although our naïve calculation might overestimate the CEF splitting, such an estimation on the order of magnitude of the CEF splitting of $10^2$ K can be trusted. Therefore, even though the CEF environment of the $Co^{2+}$ ions in NSCPO is more complicated compared with that in NBCPO due to its lower crystallographic symmetry, the ground state of NSCPO composing of the lowest-energy Kramer doublets can still be safely treated as an effective $\tilde{\mathbf{J}} = 1/2$ state at low temperatures, similar to the case in NBCPO.

Finally, the magnetic property of the polycrystalline NSCPO with the monoclinic structure has been characterized using magnetization measurements, and compared with that of polycrystalline NBCPO with the trigonal structure. As the CEF from the surrounding $CoO_6$ octahedra and the SOC can result in a Kramers doublet for the $Co^{2+}$ ions, a ground state with effective $S = 1/2$ is expected for both NBCPO and NSCPO. As shown in Fig. 5(a), the dc magnetic susceptibility ($\chi$) of NSCPO does not exhibit any feature of long-range magnetic order down to 2 K. However, its inverse ($1/\chi$) shows a change of slope around 50 K. By doing the Curie-Weiss fitting to the high-temperature ($200 < T < 300$ K) and low-temperature ($2 < T < 20$ K) parts, the effective moment of $Co^{2+}$ ions are estimated to be 5.14 $\mu_B$ and 4.10 $\mu_B$, respectively, suggesting the change of its spin from a high-temperature $S = 3/2$ state to the low-temperature effective $S = 1/2$ ground state. The Curie-Weiss temperature, $\theta_{CW}$, is determined to be -2.94 K for NSCPO from the low-temperature fitting, suggesting an antiferromagnetic interaction within the triangular Co network. Similar analysis of the dc magnetic susceptibility of polycrystalline NBCPO (Fig. 5(b)) yields the effective moments of 5.31 $\mu_B$ and 3.96

$\mu_B$ for Co$^{2+}$ ions at high and low temperatures, respectively, as well as the Curie-Weiss temperature of $\theta_{CW}$ = -2.82 K, which is well consistent with the values determined from single-crystal NBCPO. [16] Therefore, the dominant antiferromagnetic exchange energy ($J$) in monoclinic NSCPO and trigonal NBCPO is comparable, as $J$ is proportional to $\theta_{CW}$, according to the mean field theory, suggesting the robustness of magnetic exchange interaction against the Ba/Sr substitutions. However, as monoclinic NSCPO displays less geometrical frustration compared with trigonal NBCPO, a higher antiferromagnetic ordering temperature than $T_N$ ~ 148mK (for NBCPO) is expected to be observed below 2 K for NSCPO, if the Co$^{2+}$ spins order. Further measurements at extremely low temperatures for NSCPO in both polycrystalline and single-crystal forms are still ongoing to clarify its magnetic ground state for comparison with the QSL candidate NBCPO, which is helpful for tuning the intriguing magnetic interactions to realize exotic quantum states in the family of such spin-1/2 triangular-lattice antiferromagnets. (Fig. 7).

**CONCLUSION**

In conclusion, using x-ray and neutron powder diffraction as complementary methods, we have investigated the structural evolution of Na$_2$Ba$_{1-x}$Sr$_x$Co(PO$_4$)$_2$ (NBSCPO), a spin-1/2 triangular-lattice antiferromagnet as a quantum spin liquid candidate, with the chemical doping of Sr. The trigonal and monoclinic phase are found to be structurally stable at low ($x \leq 0.3$) and high ($x \geq 0.6$) Sr doping levels, and they coexist at intermediate doping levels. The doping-induced trigonal-monoclinic structural transformation is accompanied by slight rotations of CoO$_6$ octahedra and sizeable displacements of Ba$^{2+}$/Sr$^{2+}$ ions, leading to a lattice contraction and accordingly a decrease of both interlayer and intralayer Co-Co distances. Such a doping-induced structural transformation in NBSCPO is supported by our first-principles calculations and Raman spectroscopy. The dc magnetic susceptibility measurements reveal a comparable antiferromagnetic exchange energy for the end members, NSCPO and NBCPO. The monoclinically-distorted NSCPO does not show any long-range magnetic order down to 2 K and a similar negative Curie-Weiss temperature as $Na_2BaCo(PO_4)_2$ with a perfect triangular lattice, suggesting the robustness of magnetic exchange interaction against the Ba/Sr substitutions.

**Acknowledgments**

This work is financially support by the National Natural Science Foundation of China (Grant No. 12074023, NO. 12074024), and the Fundamental Research Funds for the Central Universities in China. The authors are grateful to Wei Li for valuable discussions, and Zirong Ye for the technical assistance in the MPMS measurements.

∗ These authors contributed equally to this work.

† Electronic address: lhhe@iphy.ac.cn

‡ Electronic address: wtjin@buaa.edu.cn

Table 1: Refinement results of the structural parameters of Na$_2$SrCo(PO$_4$)$_2$ (NSCPO) at room temperature.

| | | Space group: $P2_1/a$ | | | | |
|---|---|---|---|---|---|---|
| | | $a = 9.1940(1)$Å | $b = 5.2593(1)$Å | $c = 13.5313(2)$Å | | |
| | | $\alpha = 90°$ | $\beta = 90.069(1)°$ | $\gamma = 90°$ | | |
| Atom | Site | $x$ | $y$ | $z$ | Occupancy | $B_{iso}$ (Å) |
| Na$_1$ | 4e | 0.185(1) | 0.520(2) | 0.908(1) | 1 | 1.04(4) |
| Na$_2$ | 4e | 0.149(1) | 0.515(3) | 0.411(1) | 1 | 2.07(4) |
| Sr | 4e | 0.035(1) | 0.044(1) | 0.750(1) | 0.94(1) | 1.00(1) |
| Co$_1$ | 2d | 0.000 | 0.000 | 0.500 | 0.5 | 0.97(9) |
| Co$_2$ | 2a | 0.000 | 0.000 | 0.000 | 0.5 | 2.19(9) |
| P$_1$ | 4e | 0.178(1) | 0.526(2) | 0.633(1) | 1 | 1.16(2) |
| P$_2$ | 4e | 0.148(1) | 0.488(2) | 0.131(1) | 1 | 0.91(2) |
| O$_1$ | 4e | 0.146(1) | 0.521(2) | 0.743(1) | 1 | 0.24(1) |
| O$_2$ | 4e | 0.345(1) | 0.514(2) | 0.614(1) | 1 | 0.65(2) |
| O$_3$ | 4e | 0100(1) | 0.299(2) | 0.583(1) | 1 | 0.96(2) |
| O$_4$ | 4e | 0.132(1) | 0.782(2) | 0.588(1) | 1 | 0.71(2) |
| O$_5$ | 4e | 0.178(1) | 0.542(1) | 0.241(1) | 1 | 0.94(2) |
| O$_6$ | 4e | 0.299(1) | 0.497(2) | 0.078(1) | 1 | 0.87(2) |
| O$_7$ | 4e | 0.081(1) | 0.226(2) | 0.114(1) | 1 | 0.52(2) |
| O$_8$ | 4e | 0.049(1) | 0.695(2) | 0.089(1) | 1 | 0.97(2) |
| Goodness of NPD fitting: | | $R_p = 2.75$ | | $R_{wp} = 3.45$ | | $\chi^2 = 2.04$ |
| Goodness of XRD fitting: | | $R_p = 4.95$ | | $R_{wp} = 7.10$ | | $\chi^2 = 3.90$ |

Table 2: Comparisons of the calculated free energies between different possible structural phases of NBCPO and NSCPO. The energy has been normalized to a formula unit containing one Co$^{2+}$ ion.

| Composition | Energy of trigonal phase | Energy of monoclinic phase |
|---|---|---|
| NBCPO | 0 | 409.9 meV |
| NSCPO | 231.9 meV | 0 |

Table 3: The calculated CEF parameters of NBCPO and NSCPO.

| | CEF parameters(meV) |
|---|---|
| NBCPO | $B_2^0 = -2.326, B_4^0 = 0.425, B_4^3 = -11.262$ |
| NSCPO-Co$_1$ | $B_2^0 = 1.820, B_2^1 = 19.100, B_2^2 = -2.482, B_2^{-1} = 8.486, B_2^{-2} = -2.990$ <br> $B_4^0 = -0.572, B_4^1 = 0.003, B_4^2 = 0.060, B_4^3 = 0.960, B_4^4 = -3.163$ <br> $B_4^{-1} = 0.007, B_4^{-2} = 0.055, B_4^{-3} = -0.434, B_4^{-4} = 0.269$ |
| NSCPO-Co$_2$ | $B_2^0 = 1.595, B_2^1 = 12.252, B_2^2 = -1.442, B_2^{-1} = 7.439, B_2^{-2} = -4.331$ <br> $B_4^0 = -0.600, B_4^1 = 0.012, B_4^2 = 0.046, B_4^3 = 0.556, B_4^4 = -3.199$ <br> $B_4^{-1} = 0.006, B_4^{-2} = 0.105, B_4^{-3} = -0.388, B_4^{-4} = 0.407$ |

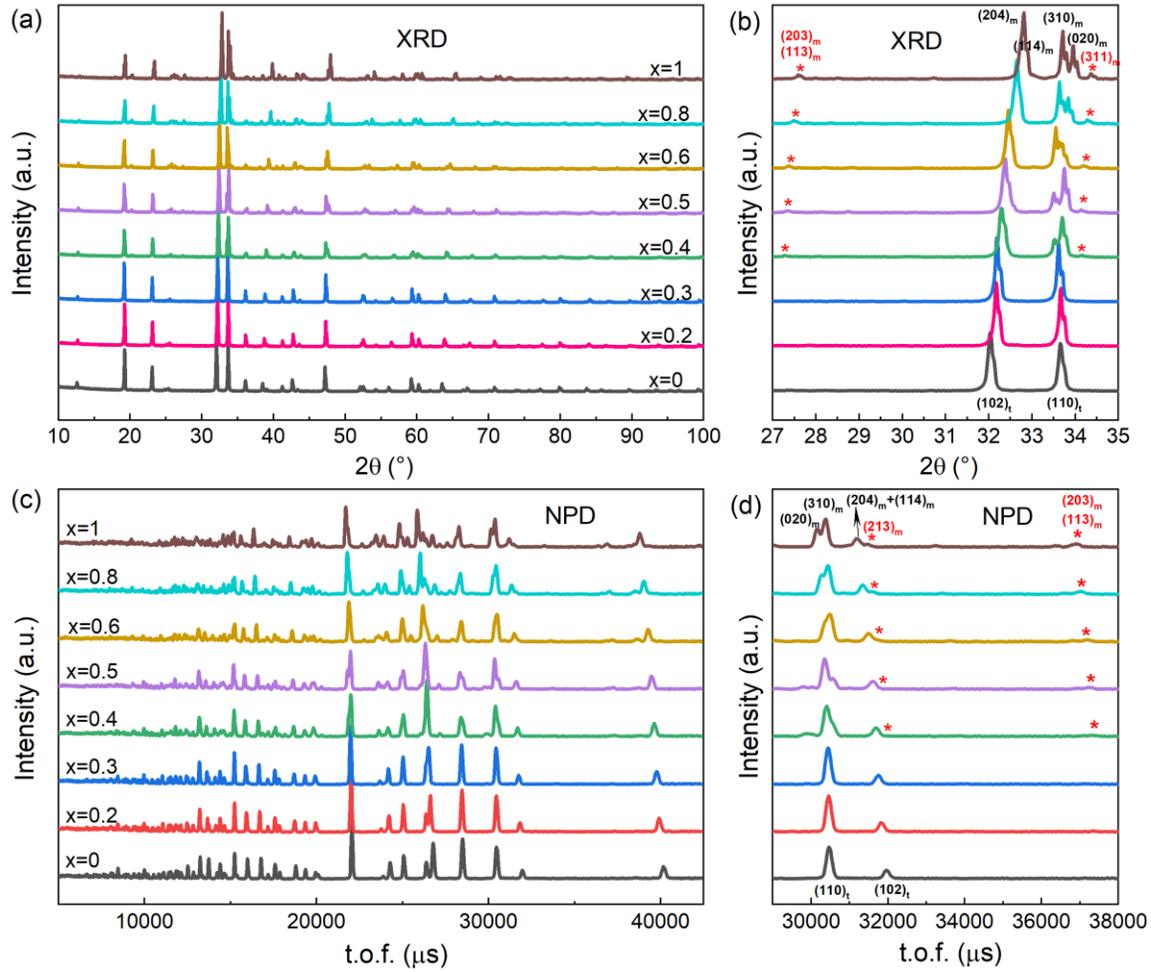

Figure 1: XRD (a) and NPD (c) patterns of $Na_2Ba_{1-x}Sr_xCo(PO_4)_2$ at room temperature for $x = 0, 0.2, 0.3, 0.4, 0.5, 0.6, 0.8$, and 1. The enlarged XRD patterns from $2\theta = 27°$ to $35°$ and the NPD patterns from t.o.f = 29000 $\mu$s to 38000 $\mu$s are plotted in (b) and (d), respectively, supporting the structural transformation from trigonal to monoclinic symmetry induced by Sr doping. The asterisks in (b) and (d) mark the emergent weak reflections allowed only in the monoclinic phase.

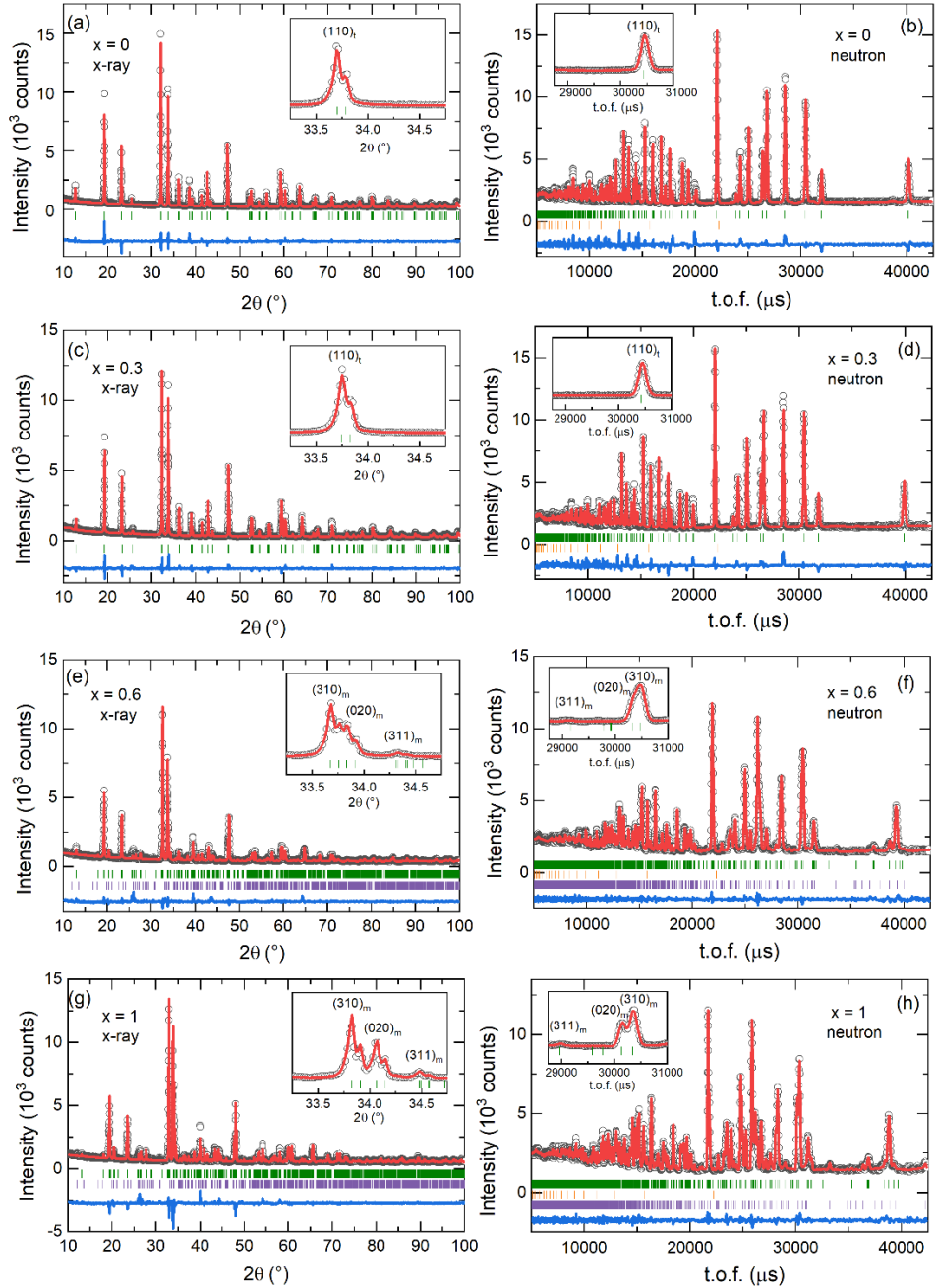

Figure 2: XRD and NPD patterns of $Na_2Ba_{1-x}Sr_xCo(PO_4)_2$ for $x = 0$ (a and b), 0.3 (c and d), 0.6 (e and f) and 1 (g and h) and corresponding Rietveld refinements. The circles represent the observed intensities, and the solid lines are the calculated patterns. The differences between the observed and calculated intensities are shown at the bottoms. The vertical bars in olive, orange, and purple colors correspond to the expected Bragg reflections from the NBSCPO main phase, vanadium sample container and $Sr_2P_2O_5$ impurity, respectively. The insets show the splitting of $(110)_t$ reflection and the emergence of weak reflection $(311)_m$ upon the trigonal-monoclinic structural transformation induced by Sr doping.

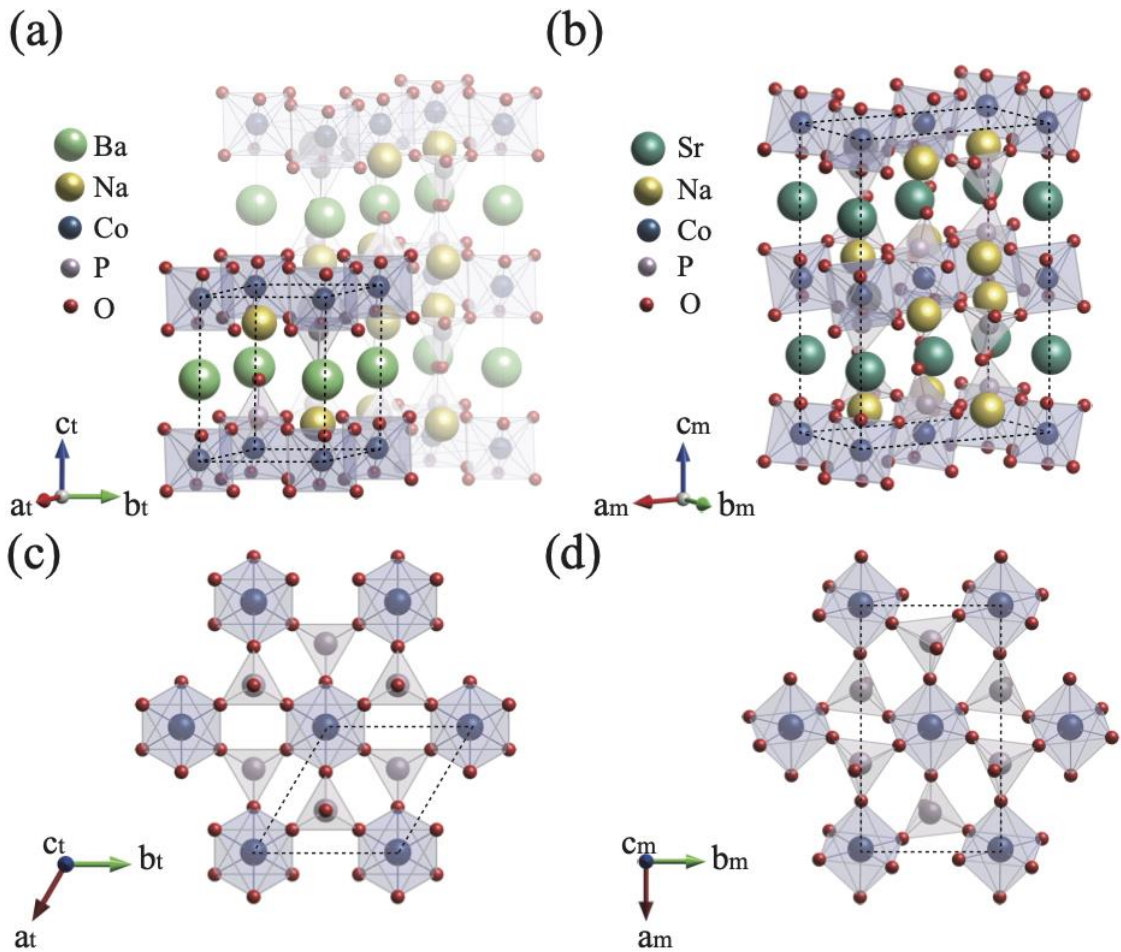

Figure 3: Crystal structures of trigonal NBCPO (a) and monoclinic NSCPO (b), with the dashed lines marking the corresponding unit cells. The in-plane structure viewed along the c axis is illustrated in (c) for NBCPO and (d) for NSCPO.

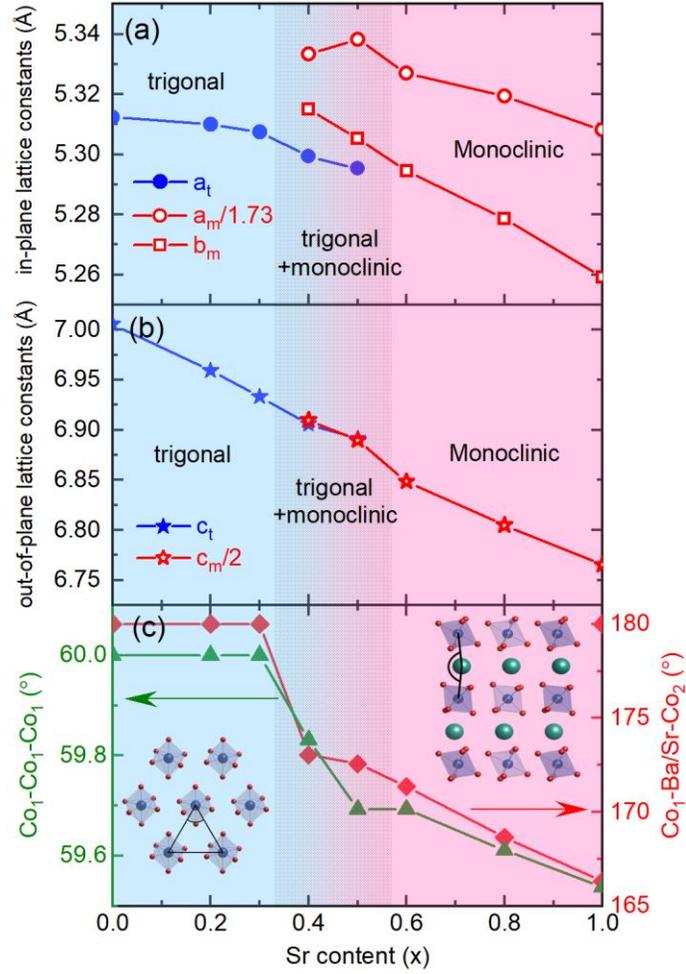

Figure 4: The in-plane (a) and our-of-plane (b) lattice constants of $Na_2Ba_{1-x}Sr_xCo(PO_4)_2$, as well as the $Co_1$-$Co_1$-$Co_1$ and $Co_1$-Ba/Sr-$Co_2$ bond angles (c) as functions of the Sr doping level. Note that the unit cell in monoclinic notation can be converted into that in trigonal notation, considering $a_m \approx \sqrt{3}a_t$, $b_m \approx b_t$, and $c_m \approx 2c_t$. The regime showing a pure trigonal or monoclinic phase is marked by blue or red color, with related structural parameters represented by solid and hollow symbols, respectively. The shaded area represents the regime with the coexistence of two structural phases at intermediate Sr doping levels.

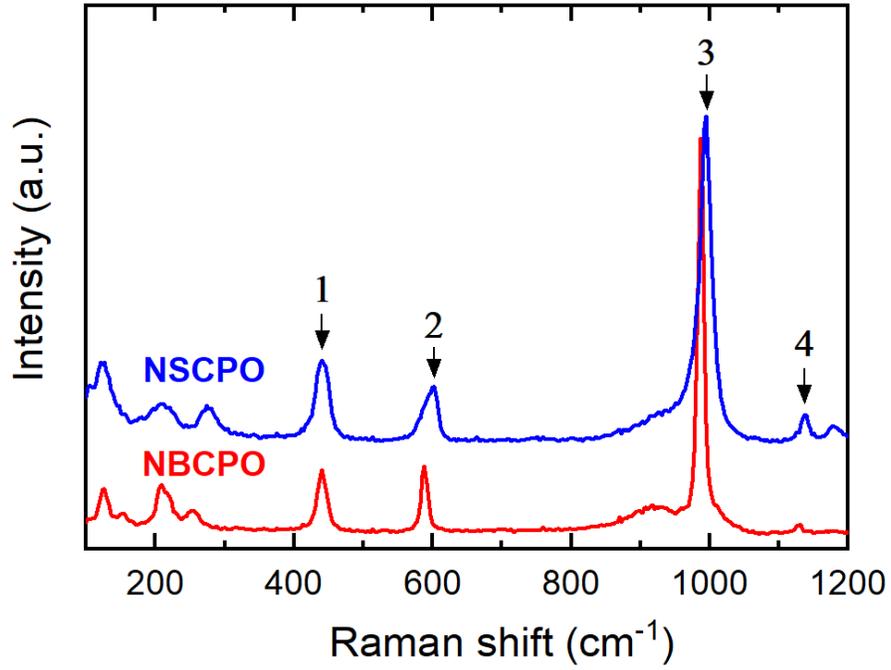

Figure 5: The Raman spectra of polycrystalline NSCPO and NBCPO collected at room temperature. The vertical arrows mark the Raman modes in the range of 400-1200 cm$^{-1}$ associated with the vibration modes of the PO$_4$ tetrahedra.

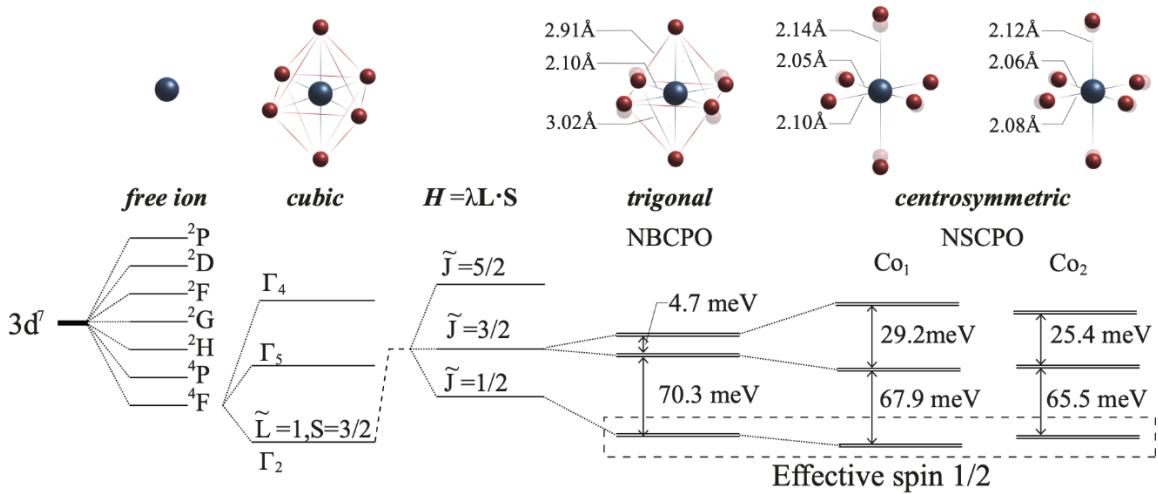

Figure 6: The diagram of energy-level alignments of $Co^{2+}$ ions. The ground state $^4F$ for a free $Co^{2+}$ ion splits into three energy levels ($\Gamma_2$, $\Gamma_5$, and $\Gamma_4$) in a cubic crystal field. In the presence of SOC, the $\Gamma_2$ state ($\tilde{L} = 1, S = 3/2$) evolves into three energy levels characterized by $\tilde{J} = 1/2$, $3/2$ and $5/2$, respectively. For trigonal NBCPO and monoclinic NSCPO, the trigonal and centrosymmetric distortion further enhances the CEF splitting, yielding an effective $\tilde{J} = 1/2$ ground state for both NBCPO and NSCPO. The blue and red spheres represent the $Co^{2+}$ and $O^{2-}$ ions, respectively, with the shadows marking the original positions of $O^{2-}$ ions in non-distorted cubic crystal field. Please note that different colors of the chemical bonds represent different bond lengths in the $CoO_6$ octahedra, as determined experimentally.

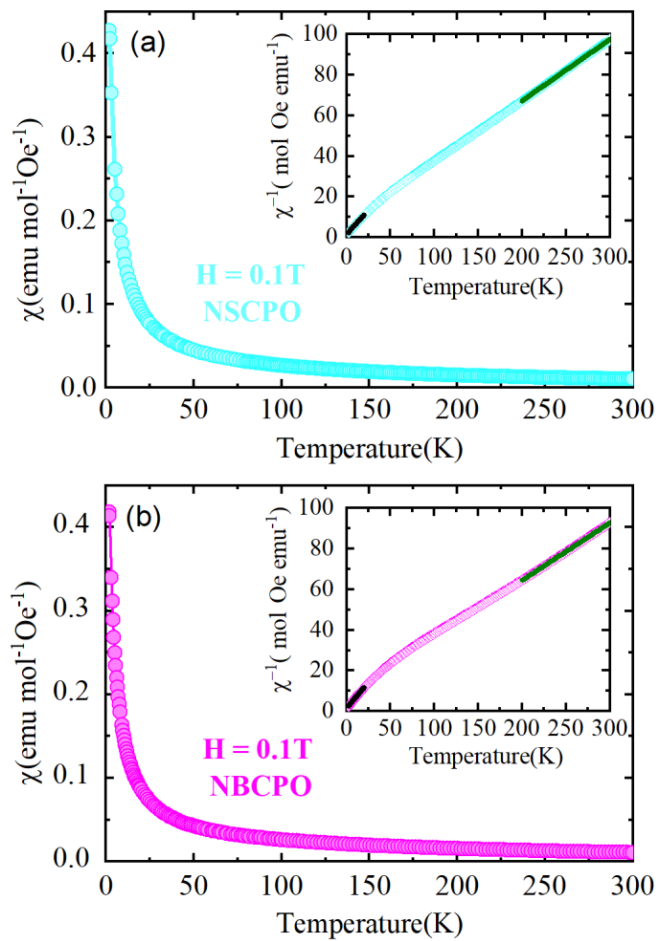

Figure 7: The dc magnetic susceptibility of polycrystalline NSCPO(a) and NBCPO (b), measured in an applied magnetic field of 0.1 T in zero-field-cooling (ZFC) mode. The insets show the inverse of the susceptibility, with the Curie-Weiss fitting to the high-temperature (200 < T < 300 K) and low-temperature (2 < T < 20 K) parts plotted as olive and black solid lines, respectively.